\documentclass[aps,prb,twocolumn,superscriptaddress]{revtex4-1}
\usepackage{amsmath,amsthm,amssymb}
\usepackage[dvipdfmx]{graphicx}
\usepackage{amsmath,bm}
\usepackage{color,ulem}
\usepackage{ifthen}

\newcount \commentout
\newcommand{\1}{\mbox{1}\hspace{-0.25em}\mbox{l}}

\newlength{\figwidth}
\newlength{\figlarge}
\begin{document}
\title{
Reduction of $\mathbb{Z}$ classification of a two-dimensional weak topological insulator
\\
- real-space DMFT study -
}
\author{Tsuneya Yoshida}
\affiliation{Department of Physics, Kyoto University, Kyoto 606-8502, Japan}
\author{Norio Kawakami}
\affiliation{Department of Physics, Kyoto University, Kyoto 606-8502, Japan}
\date{\today}
\begin{abstract}
One of the remarkable interaction effects on topological insulators is the reduction of topological classification in free-fermion systems. 
We address this issue in a bilayer honeycomb lattice model by taking into account temperature effects on the reduction. 
Our analysis, based on the real-space dynamical mean field theory, elucidates the following results. 
(i) Even when the reduction occurs, the winding number defined by the Green's function can take a nontrivial value at zero temperature. 
(ii) The winding number taking the nontrivial value becomes consistent with the absence of gapless edge modes due to Mott behaviors emerging only at the edges. 
(iii) Temperature effects can restore the gapless edge modes, provided that the energy scale of interactions is smaller than the bulk gap. In addition, we observe the topological edge Mott behavior only in some finite temperature region.
\end{abstract}
\maketitle

\section{Introduction} 
After the discovery of topological insulators, topological aspects of free-fermion systems have been extensively studied.\cite{TI_review_Hasan10,TI_review_Qi10} 
In these systems, the ground state wave function has topologically nontrivial properties which can be characterized with topological invariants calculated from the Bloch wave function.
Nontrivial topology (, or a nontrivial value of topological invariant) in the bulk predicts gapless edge modes, which is known as the bulk-edge correspondence.
The presence of these gapless edge modes is a source of characteristic magneto-electric responses.
Furthermore, gapless edge modes in the topological superconductor are described as Majorana fermions whose experimental realization has been addressed for one-dimensional quantum wires, recently.\cite{Majorana_Mourik,Majorana_Rokhinson2012,Majorana_Das2012}

In parallel with study of free-fermion systems, topological insulators for strongly correlated compounds have been proposed.\cite{NaIrO_Nagaosa09,Heusler_Chadov10,Heusler_Lin10,Takimoto_SmB6_2011,skutterudites_Yan12,Weng_YbB12_14} 
The first principles calculation has pointed out that $\mathrm{SmB}_6$ can be a strong topological insulator.\cite{Takimoto_SmB6_2011} 
Furthermore, correlation effects on topological insulators are expected to induce novel phenomena. These facts urge us to study correlation effects on topological insulators.

For analysis of topological insulators with electron interactions, a topological invariant defined in the bulk plays an important role. In non-interacting systems, this quantity can be calculated from the Bloch wave function. 
The topological invariant for free-fermion systems can be extended to interacting systems.\cite{Ishikawa_IQHE_1987,ch_num_Haldane,Volovik_textbook2009,Gurarie_2011,Essin_12,Wang12,Wang_12_prb} 
Even though the Bloch wave function is no longer well-defined in interacting systems, the single-particle Green's function is still available. 
Thus, the topological invariant can be defined with the Green's function, which means that the group structure of topological insulators (e.g., $\mathbb{Z}$ or $\mathbb{Z}_2$) is the same as that of the non-interacting case.

Interestingly, however, it is reported that the interactions can induce the reduction of topological classification of free-fermion systems.\cite{Fidkowski_1Dclassificatin_11,Turner11,YaoRyu_Z_to_Z8_2013,Qi_Z_to_Z8_2013,Lu_CS_2011,Levin_CS_2012,Hsieh_CS_CPT_2014,Wang_classification,Cenke2014,Morimoto_16} 
Namely, at the non-interacting level, topological insulators/superconductors labeled by integers show gapless edge modes due to nontrivial properties, while gapless edge modes in some of these topological insulators/superconductors can be gapped out without symmetry breaking under electron interactions. 
This means that some of topological insulators/superconductors at the non-interacting level become topologically trivial in the presence of interactions, resulting in the reduction of topological classification.
The reduction of topological classification has been mainly analyzed in terms of symmetry protection of edge modes at zero temperature.

Now, one may ask the following three questions. 
(i) How the topological invariant defined by the Green's function behaves if the reduction of topological classification occurs? Does it still take a nontrivial value? 
(ii) If the topological invariant of the Green's function is well-defined and takes a nontrivial value, how this nontrivial value in the bulk becomes consistent in the absence of the gapless edge modes? 
(iii) What are temperature effects on the reduction of topological classification? 
The analysis at finite temperatures is desired because all of experiments are carried out there. 

In this paper, we address these three questions by analyzing a correlated weak topological insulator in two dimensions. 
The topological structure is characterized by the winding number at the non-interacting level, which is also well-defined even in correlated systems.
Our analysis based on the real-space dynamical mean field theory with continuous-time quantum Monte Carlo (R-DMFT+CTQMC) elucidates the following facts. 
First, the winding number can take a nontrivial value even when the reduction of topological classification occurs. 
Second, the system shows a Mott behavior (i.e., the divergence of the self-energy) only around the edge, signaling the reduction of topological classification. 
Third, temperature effects result in a gradual crossover in gapless edge modes.
Namely, the behavior of edge modes changes with increasing temperature from zero temperature: starting from the absence of gapless excitations, gapless spin excitations emerge at slightly higher temperatures, which finally leads to single-particle gapless excitations in the higher temperature region.

The rest of this paper is organized as follows. 
The next section (Sec.~\ref{sec: model_and_method}) is devoted to the topological invariant with the Green's function as well as description of our model and method. 
In Sec.~\ref{sec: results}, we show that interactions induce the reduction of topological classification from $\mathbb{Z}$ to $\mathbb{Z}_4$ and analyze the resulting properties in detail by using the R-DMFT. 
Short summary is given in Sec.~\ref{sec. summary}.

\section{Model and method}\label{sec: model_and_method}
\subsection{Lattice model}\label{sec: model}
\begin{figure}[!h]
\begin{center}
\includegraphics[width=\hsize,clip]{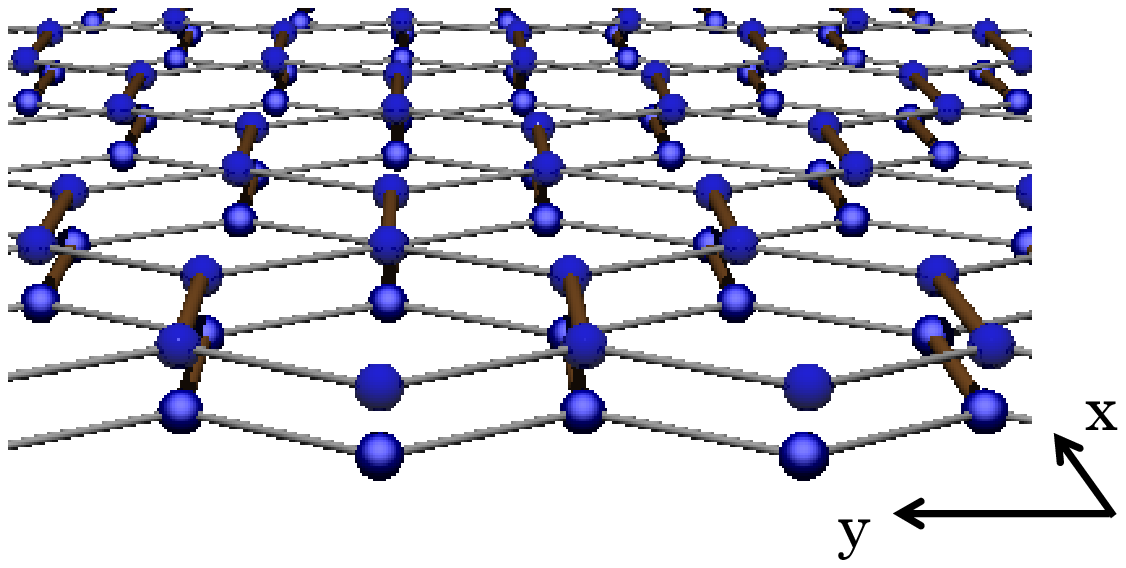}
\end{center}
\caption{(Color Online). 
Sketch of the model. 
We consider the system of an A-A stacked honeycomb lattice under the open (periodic) boundary condition for $x$- ($y$-) direction, respectively.
The parameter $t$ ($rt$) denotes hopping of electron between sites connected with a brown (gray) line. 
}
\label{fig:model}
\end{figure}

We study a bilayer honeycomb lattice model. The Hamiltonian reads
\begin{subequations}
\label{eq: model}
\begin{eqnarray}
H&=&H_0+H_\mathrm{int}, \\
H_0&=&\sum_{\langle i,j\rangle,\alpha}t_{i,j}c^\dagger_{i,\alpha,\sigma}c_{j,\alpha,\sigma},  \label{eq: model_b} \\ 
H_\mathrm{int}&=&U\sum_{i\alpha} n_{i,\alpha,\uparrow}n_{i,\alpha,\downarrow} +J\sum_{i}\bm{S}_{i,a}\cdot\bm{S}_{i,b},
\end{eqnarray}
\end{subequations}
where the operator $c^\dagger_{i,\alpha,\sigma}$ creates an electron at site $i$ and in layer $\alpha=a,b$ and spin $\sigma=\uparrow,\downarrow$ state. 
$t_{i,j}$ is intra-layer hopping between sites $i$ and $j$, and the inter-layer hopping is assumed to be zero. 
If an electron hops in $x$-direction, we set $t_{i,j}=t$, otherwise $t_{i,j}=rt$ ($r$ is a real number).
A sketch of the hopping $t_{i,j}$ is shown in Fig.~\ref{fig:model}; the brown (gray) bonds denote hopping with $t$ ($rt$) respectively.
The electrons interact via intra-layer on-site interaction $U$ and spin-exchange interaction $J$.
We consider the system under the open (periodic) boundary condition for $x$- ($y$-) direction respectively. 
The system is composed of $L$ rings along the $y$-direction. $i_x(=0,\cdots,L-1)$ denotes the $x$-coordinate of these rings.

\subsection{Numerical approach: R-DMFT}\label{subsec: r-DMFT}
For systematic analysis of the bulk and the edge properties, we consider the system with a cylinder geometry. Namely, we impose the open (periodic) boundary condition for $x$- ($y$-) direction, respectively.
This system is an inhomogeneous correlated system.
The real-space dynamical mean field theory,\cite{RDMFT_Potthoff,RDMFT_Okamoto,RDMFT_Rosch2008,RDMFT_Snoek} an extended version of the DMFT,\cite{DMFT_Muller_1989,DMFT_Metzner_PRL1989,DMFT_Georges_1996} enables us to treat it.

In this method, the problem is solved self-consistently. 
A brief outline of this approach is given in the following. 
In the first step, a one-dimensional subsystem of $i_x$-th sites along the $y$-axis is mapped onto an effective impurity model characterized by the cavity Green's function $\mathcal{G}_{i_x\alpha\sigma}(i\omega)$.
In the second step, applying an impurity solver to this effective model yields the self-energy $\Sigma_{i_x\alpha\sigma}(i\omega)$.
In the third step, the effective models are updated with obtained self-energy, and then we go back to the first step. The self-consistent equation is given by
\begin{subequations}
\begin{eqnarray}
\hat{\mathcal{G}}^{-1}_{\alpha\sigma}(i\omega_n) &=&[\sum_{k_y}\hat{G}_{\alpha\sigma}(k_y,i\omega)]^{-1}+\hat{\Sigma}_{\alpha \sigma}(i\omega_n), \\
\hat{G}_{\alpha\sigma}(k_y,i\omega)&=&\{i\omega_n \1 -\hat{h}_{\alpha\sigma}(k_y)  -\hat{\Sigma}_{\alpha \sigma}(i\omega_n)\}^{-1}, 
\end{eqnarray}
with diagonal matrices
\begin{eqnarray}
\!\!\!\hat{\Sigma}_{\alpha \sigma}(i\omega_n)&:=&\mathrm{diag}\left(\Sigma_{0\alpha\sigma},\cdots, \Sigma_{i_x\alpha\sigma} ,\cdots,\Sigma_{L-1\alpha\sigma}\right), \\
\!\!\!\hat{\mathcal{G}}_{\alpha\sigma}(i\omega_n)&:=&\mathrm{diag}\left(\mathcal{G}_{0\alpha\sigma}, \cdots, \mathcal{G}_{i_x \alpha\sigma}, \cdots, \mathcal{G}_{L-1\alpha\sigma}\right),
\end{eqnarray}
where $\hat{h}_{\alpha\sigma}(k_y)$ is the Fourier transform of the hopping matrix and $\hat{G}_{\alpha\sigma}(k_y,i\omega)$ is the Green's function for the lattice model.
\end{subequations}
The effective impurity model is essentially the same as that of two-orbital systems.
With path integral formalism, the effective model for $i_x$-th sites are written as
\begin{subequations}
\begin{eqnarray}
\mathcal{Z}_{\mathrm{imp},i_x}&=&\int \Pi_{\sigma} \mathcal{D}\hat{\Psi}_{i_x,\sigma}\mathcal{D}\hat{\bar{\Psi}}_{i_x,\sigma} e^{-S_\mathrm{imp}}, \\
\mathcal{S}_{\mathrm{imp},i_x}&=&-\int d\tau d\tau' \left[ \sum_{\sigma,\alpha} \bar{c}_{i_x,\alpha,\sigma}(\tau) \mathcal{G}^{-1}_{i_x,\alpha}(\tau-\tau') c_{i_x,\alpha,\sigma}(\tau') \right] \nonumber\\
&& \quad\quad\quad\quad\quad  + \int d\tau H_{\mathrm{imp},i_x}, \\
H_{\mathrm{imp},i_x} &=& U\sum_{\alpha=a,b} (n_{i_x,\alpha,\uparrow}-\frac{1}{2}) (n_{i_x,\alpha,\downarrow}-\frac{1}{2}) +J\bm{S}_{i_x,a}\cdot \bm{S}_{i_x,b}, \nonumber \\
\end{eqnarray}
\end{subequations}
with $\alpha=a,b$.
To solve the effective model we employ the continuous-time quantum Monte Calro\cite{CTQMC_Rubtsov,Werner_CTQMC_PRL2006,Werner_CTQMC_CTQMC2006,Haule_CTQMC_CTQMC2007,Gull_RMP2011} which is a powerful tool to analyze multi-orbital systems.

\subsection{The winding number}\label{subsec: winding_GF}
For $r<0.5$ and $U=J=0$, the model (\ref{eq: model}) having two-sublattices A and B realizes a weak topological insulator which can be regarded as an array of one-dimensional topological insulators with chiral symmetry along $y$-direction in the momentum space. Accordingly, gapless edge modes are observed along zigzag edges, while no gapless edge mode is observed along armchair edges. 
The topological properties of this insulator are protected by the chiral symmetry and characterized by the winding number which is defined as
\cite{Gurarie_2011,Essin_12,Wang_12_prb,Manmana_12}
\begin{eqnarray}
\nu&=& \int^{2\pi}_0 \frac{dk_x}{4\pi i} \mathrm{Tr}[\tau_3 \hat{g}^{-1}(k_x)\partial_{k_x} \hat{g}(k_x)], 
\label{eq: nu}
\end{eqnarray}
where $g_{n m}(k_x)$  with $n, m\in A, B$ is the full Green's function for $i\omega=0$ and $k_y=0$ under the torus geometry. In our bilayer system, $g_{n m}$ is a $4\times 4$-matrix.
$\tau$'s are Pauli matrices acting on the sublattice space. 
The winding number is integral  (i.e., $\nu \in \mathbb{Z}$), which reflects the mathematical fact, $\pi_1[U(2m)/U(m)\times U(m)]=\mathbb{Z}$ with a positive integer $m$.\cite{Essin_12}

\section{Results}\label{sec: results}
First, we address the instability of gapless edge modes against interactions, which leads to the reduction of the topological classification from $\mathbb{Z}$ to $\mathbb{Z}_4$.
After that, we discuss the numerical results obtained by the R-DMFT+CTQMC.
\subsection{$\mathbb{Z}_4$ classification}\label{subsec: breakdown}
We show that gapless edge modes are destroyed in four copies of a weak topological insulator in two dimensions with chiral symmetry. 
This result indicates that the topological classification $\mathbb{Z}$ in free fermions changes to $\mathbb{Z}_4$. 
We observe the instability of gapless edge modes against the interactions with the following two steps: (i)~Adiabatically deforming the effective model of Eq.~(\ref{eq: model_b}) into a simple model; (ii)~Introducing the interactions to destroy the gapless edge modes without symmetry breaking.

Let us start with the effective Hamiltonian $\mathcal{H}$ of the system under the cylinder geometry,
\begin{eqnarray}
\mathcal{H}&=&\int dx dk_y \left[it(k_y)\partial_x \tau_2+m(x,k_y)\tau_1 \right]\otimes\1_{4\times4},
\end{eqnarray}
where $t(k_y)$ describes hopping of electrons with the momentum $k_y$ along the $x$-direction. 
$\1_{4\times4}$ denotes the four-dimensional identity matrix acting on the spin and the pseudospin space of $a$ and $b$.
In this basis the operator for chiral symmetry is given by $\tau_3$; $\tau_3 \mathcal{H} \tau_3=-\mathcal{H}$.
Zero modes are localized at the kink for arbitrary $k_y$, where $m(x,k_y)$ changes its sign. 
Thus, we can adiabatically deform the Hamiltonian, $\mathcal{H}$, into the following Hamiltonian 
\begin{eqnarray}
\mathcal{H}'&=&\int dx dk_y \left[ i\partial_x \tau_2+m_0\mathrm{sign}(x)\tau_1 \right]\otimes\1_{4\times4},  \nonumber \\
 &=&\int dx dy \left[ i\partial_x \tau_2+m_0\mathrm{sign}(x)\tau_1 \right]\otimes\1_{4\times4},
\label{eq: effective Hami}
\end{eqnarray}
Electrons in this deformed model, Eq.~(\ref{eq: effective Hami}), do not hop in the $y$-direction. So, each decoupled chain along the $x$-direction behaves as an ordinary one-dimensional topological insulator with chiral symmetry.

In this one-dimensional topological insulator, we can observe the instability of gapless edge modes in the following way:\cite{Z_to_Zn_Fidkowski_10} 
(a) Introducing $Un_{i_x,\alpha,\uparrow}(x)n_{i_x,\alpha,\downarrow}(x)$ $\alpha=a,b$ reduces the system to the effective spin model showing a free spin of $S=1/2$ at the edge of each chain; (b) Introducing antiferromagnetic coupling $J\bm{S}_{i_x,a}\cdot\bm{S}_{i_x,b}$ gaps out these free spins.

Therefore we can gap out all of the gapless edge modes without symmetry breaking, which indicates the $\mathbb{Z}_4$ classification.

\subsection{Numerical results}\label{sec: n_results}
Here, we address the reduction of the topological classification by using the R-DMFT.
We focus on the paramagnetic phase because no continuous symmetry is broken in two-dimensional systems at finite temperatures.
Here, we set the parameter $L=40$ and $r=0.2$. In the following we mainly focus on the case of $U=J$. This is because the qualitative behaviors do not differ from other cases of interaction strength except for the temperature effects which we will discuss in Sec.~\ref{subsec: temperature effects}.

\begin{figure}[!h]
\begin{center}
\includegraphics[width=\hsize,clip]{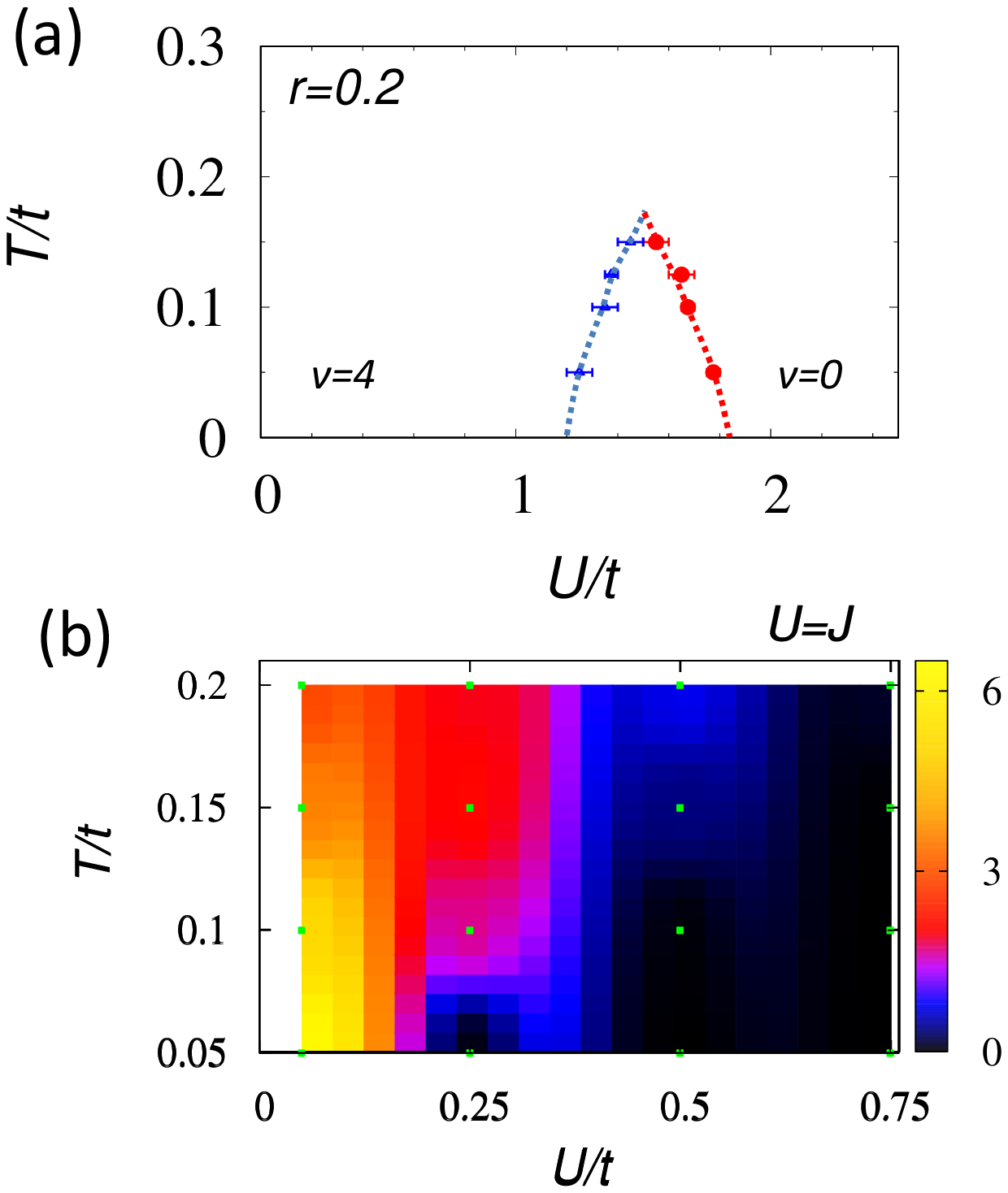}
\end{center}
\caption{(Color Online). 
Phase diagrams for interaction $U$ vs. temperature $T$ with $U=J$. 
(a) The dashed line denotes the region where hysteresis behaviors are observed in the bulk and the edge.
In the colored region, the Mott behavior is observed only around the edges.
(b) The colors denote the spectral weight of single-electron excitations at zero energy along the edge, $i_x=0$.
}
\label{fig: phase}
\end{figure}

The main results are summarized in Fig.~\ref{fig: phase},  which are obtained under the cylinder geometry. 
First, we elucidate how the winding number behaves when the system shows the reduction of the topological classification for free fermions. 
As seen in Fig.~\ref{fig: phase}(a), switching on the interactions $U$ and $J$ does not change the winding number in the bulk; the winding number remains four for sufficiently weak interactions. 
We note that the gapless edge modes are destroyed in the region of $0.25t\lesssim U$ as seen in Fig.~\ref{fig: phase}(b) where the local density of states at zero energy and $i_x=0$ is plotted.

Second, we elucidate how the winding number taking the nontrivial value becomes consistent with the absence of gapless edge modes. 
We demonstrate that the system shows Mott behaviors only around the edges, making the winding number and the absence of gapless modes consistent. 
The Mott behaviors only around the edge are observed via a divergence of the self-energy and an abrupt change of double occupancy.

Finally, we elucidate that the region of $\nu=4$ shows difference from the trivial band insulator in the finite temperature region, although these two phases are topologically identical at zero temperature.

In the following, we explain the details.
In Sec.~\ref{subsec: results_breakdown}, we analyze the reduction of the topological classification in the bulk and the edges and elucidates the behaviors of the winding number.
In Sec.~\ref{subsec: temperature effects}, we address the temperature effects on the reduction of the classification.
\subsubsection{The reduction and the winding number}\label{subsec: results_breakdown}

\begin{figure}[!h]
\includegraphics[width=85mm]{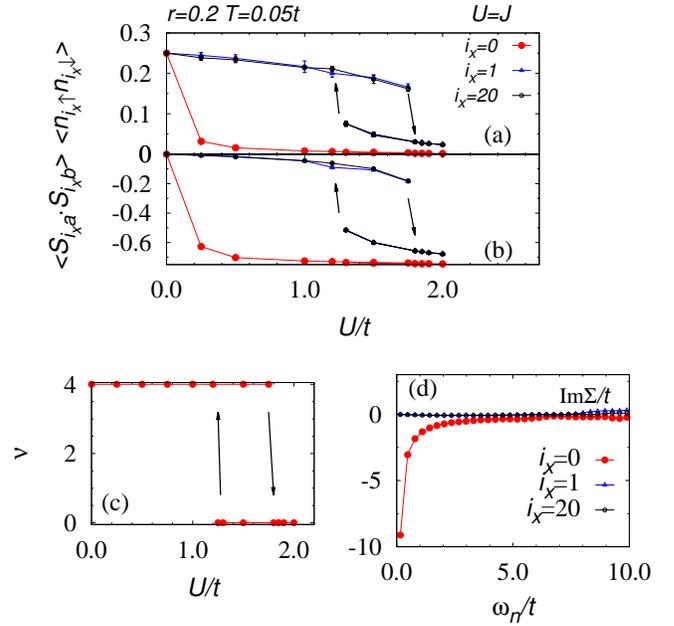}
\caption{(Color Online). 
(a) and (b) Interaction dependence of double occupancy and spin correlations respectively.
(c) Winding number as a function of interaction strength. The winding number is obtained from the single-particle Green's function at $i_x=20$ for $T=0.05t$.
(d) The imaginary part of the self-energy at each site for $U=J=t$.
}
\label{fig:docc_winding}
\end{figure}

In this section, we show that the winding number takes four, while the gapless edge modes are destroyed due to correlations.
We then elucidate how the nontrivial value of the winding number becomes consistent with the absence of gapless edge modes.
For this aim, we focus on the low temperature region $T=0.05t$, where the bulk electron gap is large enough than the temperature scale so that we can characterize the topological structure.

First, we discuss behaviors of local correlations as a function of interaction strength $U(=J)$.
With turning on the interactions, the double occupancy for the bulk ($i_x=20$) gradually changes [see Fig.~\ref{fig:docc_winding}(a)]. 
This behavior indicates that the bulk behaves as a renormalized band insulator. 
Correspondingly, the winding number remains four for $U \lesssim 1.25t$ [see Fig.~\ref{fig:docc_winding}(c)].
Further increasing the interactions, the system changes into the Mott insulating phase where the electrons are localized and form a dimer phase [see Fig.~\ref{fig:docc_winding}(a) and (b)]. 
Therefore, the system is no longer a renormalized band insulator, and the topological nontrivial structure is destroyed. 
We note that our results indicate a first order Mott transition because the hysteresis behavior is observed in the region of $1.25t \lesssim U \lesssim 1.775t $ for $T=0.05$.\cite{footnote_first}
Increasing temperature narrows the region of hysteresis behavior which is sandwiched by dashed blue and red lines in Fig.~\ref{fig: phase}(a). 
This indicates that the first order transition changes to a crossover in high temperature region.
In the above we have seen that if the interactions, $U=J$, are weak, then, the system behaves as a renormalized band insulator showing the winding number $\nu=4$.

\begin{figure}[!h]
\begin{center}
\includegraphics[width=65mm,clip]{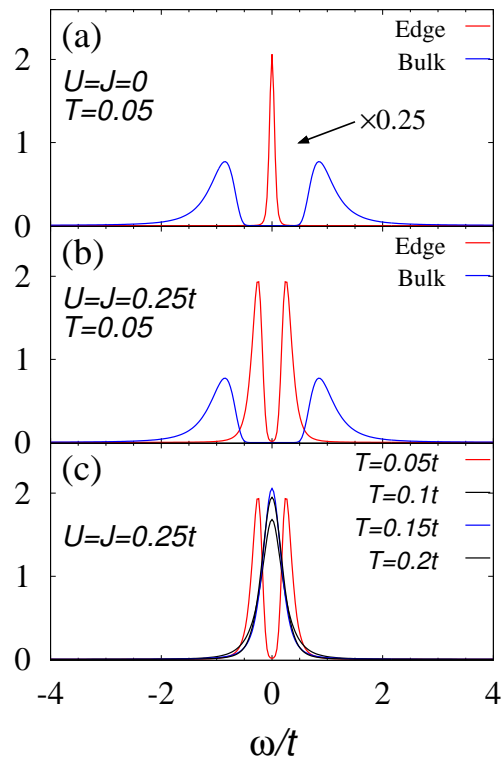}
\end{center}
\caption{(Color Online). 
(a) and (b): LDOS for the edge ($i_x=0$) and the bulk ($i_x=20$). (a) [(b)] denotes the data for $U=J=0$ and ($U=J=0.25t$), respectively. 
(c) the LDOS at the edge for $U=J=0.25t$ and several values of temperature.
}
\label{fig:LDOS for UeqJ}
\end{figure}

Here, we show that even for weak interactions, the gapless edge modes are destroyed in spite of the nontrivial values of the winding number $\nu=4$.
Destruction of gapless electron excitations at edges can be seen in Figs.~\ref{fig:LDOS for UeqJ} (a) and (b). 
In these figures, the local density of states (LDOSs), $A_{i_x}(\omega)=-\frac{1}{\pi}\mathrm{Im}G_{i_x,a,\uparrow}(\omega)$ with $G_{i_x,a,\uparrow}(\omega)=\frac{1}{2\pi}\sum_{k_y}G_{i_x,a,\uparrow}(k_y,\omega)$, for the bulk $i_x=20$ and edges $i_x=0$ are plotted. 
For the non-interacting case [Fig.~\ref{fig:LDOS for UeqJ}(a)], the gap is observed in the bulk, while a sharp peak of LDOS is observed for $\omega=0$ at the edge, signaling the presence of gapless edge modes.
Switching on the interaction does not qualitatively change the bulk gap. At the edge, however, the sharp peak at $\omega=0$ disappears [Fig.~\ref{fig:LDOS for UeqJ}(b)].
A similar behavior is also observed in the collective excitations. 
Therefore, we can conclude that the introducing interactions with $U=J$ destroys the gapless edge modes observed in the non-interacting case. 
This indicates the reduction of the topological classification for free fermions.

In the above we have seen that the winding number defined in Eq.~(\ref{eq: nu}) takes four although the system does not show any gapless modes.
At first glance, the nontrivial winding number in the bulk seems to contradict the absence of gapless edge modes. 
Here, we elucidate how these two results become consistent with each other.
Mathematically, the system does not change its topological properties under continuous deformation, provided that the Green's function at $i\omega=0$ ($\hat{g}$) does not satisfy $\det[\hat{g}^{-1}]=0$ and $\det[\hat{g}]= 0$.
Thus, when the bulk has the nontrivial topology, the following two scenarios are expected at the edge, which separates two distinct phases with different winding numbers.
(a) One possibility is that the Green's function diverges at edges and becomes singular because the topological invariant is not well-defined at edges. 
The physical meaning of this divergence is the emergence of zero energy excitations. This is indeed observed in the non-interacting case [see Fig.~\ref{fig:LDOS for UeqJ}(a)], which is consistent with the bulk-edge correspondence in the non-interacting case.
(b) The other possibility is that the Green function becomes zero (i.e., the divergence of the self-energy). 
The zeros of the Green's function are induced only by divergence of the self-energy (i.e., the zeros are absent for the non-interacting case). 
In the region of $\nu=4$ and $T=0.05t$ in Fig.~\ref{fig: phase}(a), the scenario (b) holds. In the following, we demonstrate that the system shows the Mott behavior only around the edge, which keeps the winding number $\nu=4$ and thus makes the absence of gapless edge modes consistent.
A signal of the Mott behavior emerging only around the edge is observed in the interaction dependence of the local correlations [Figs.~\ref{fig:docc_winding}(a) and (b)]. 
Once the interactions are turned on, the double occupancy and the spin correlation suddenly drop only at $i_x=0$, while these local correlations gradually change in the bulk.
In order to observe the Mott behavior more clearly, we plot the imaginary part of self-energy at $i_x=0$, $i_x=1$, and $i_x=20$. This figure indicates that the self-energy diverges at $i_x=0$ but  not at $i_x=1$ and $i_x=20$.
Therefore, we conclude that the system shows the Mott behavior only around the edge, whereby the winding number $\nu=4$ and the absence of gapless edge modes become consistent.

\subsubsection{Temperature effects}\label{subsec: temperature effects}
In this section, we address the third question; \textit{What are temperature effects on the reduction of topological classification?}
In the previous section, we have discussed the reduction. The point is that if the reduction occurs the zeros of Green's function emerge at the edge due to consistency with the nontrivial values of winding number. 
Therefore, there is a chance to recover gapless modes; if the singularity of the self-energy is suppressed, the gapless mode would show up again.
Finite temperature effects indeed recover the gapless modes.

For $U=J$, we can observe that increasing temperature suppresses the singularity of the self-energy and restores the gapless electron excitations; Fig.~\ref{fig:LDOS for UeqJ}(c) indicates that the gapless edge modes are restored with increasing temperature, while the edge modes are destroyed for $T=0.05t$.
Accordingly, the double occupancy at the edge gradually increases with increasing temperature.
Furthermore, for $U>J$, the temperature effects induce richer properties. 
Increasing temperature changes statistical properties of gapless excitations if the energy scale of interactions is smaller than the bulk gap. Namely, with increasing temperature, the bosonic excitations are restored first, after that, the fermionic excitations are restored.
Suppose that the edge modes are gapped out by interactions in the phase with the winding number $\nu=4$.
Then, with increasing temperature a topological edge Mott state emerges, which is accompanied by gapless edge modes only in the spin excitations. 
This is because the phase with $\nu=4$ is decoupled to two-copies of the weak topological insulator with $\nu=2$, each of which has gapless spin excitations as edge modes, if temperature effects are dominant than the spin exchange interactions but smaller than the on-site interaction $U$.

\begin{figure}[!h]
\begin{minipage}{1\hsize}
\begin{center}
\includegraphics[width=\hsize,clip]{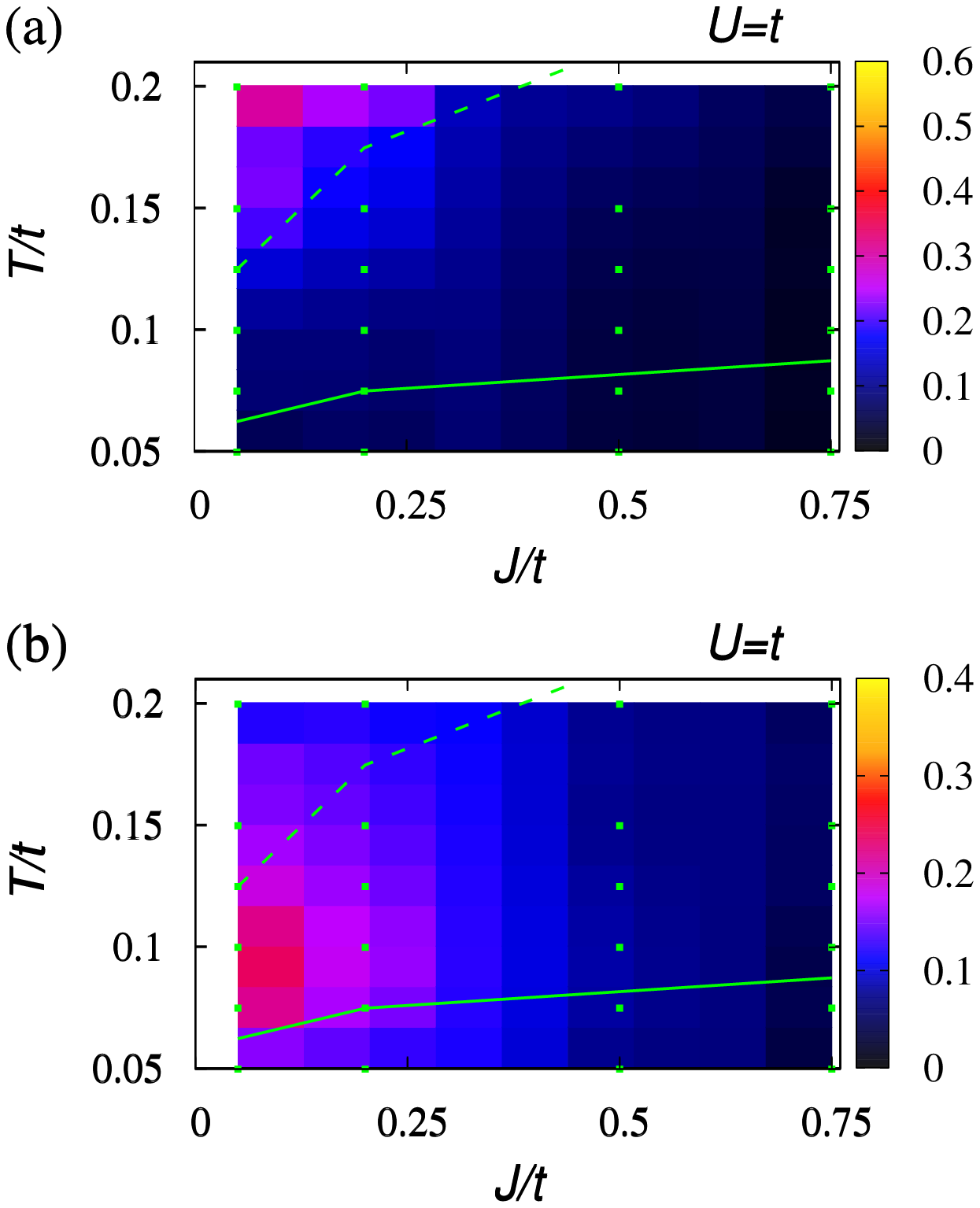}
\end{center}
\end{minipage}
\begin{minipage}{0.45\hsize}
\begin{center}
\includegraphics[width=50mm,clip]{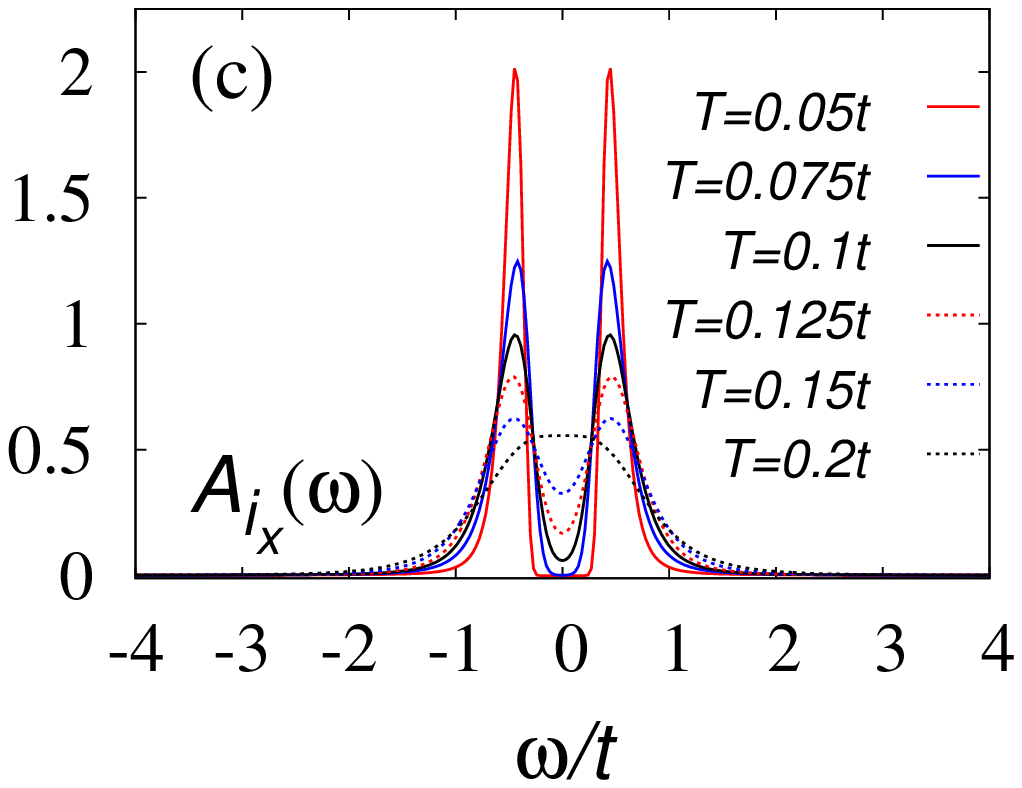}
\end{center}
\end{minipage}
\begin{minipage}{0.45\hsize}
\begin{center}
\includegraphics[width=50mm,clip]{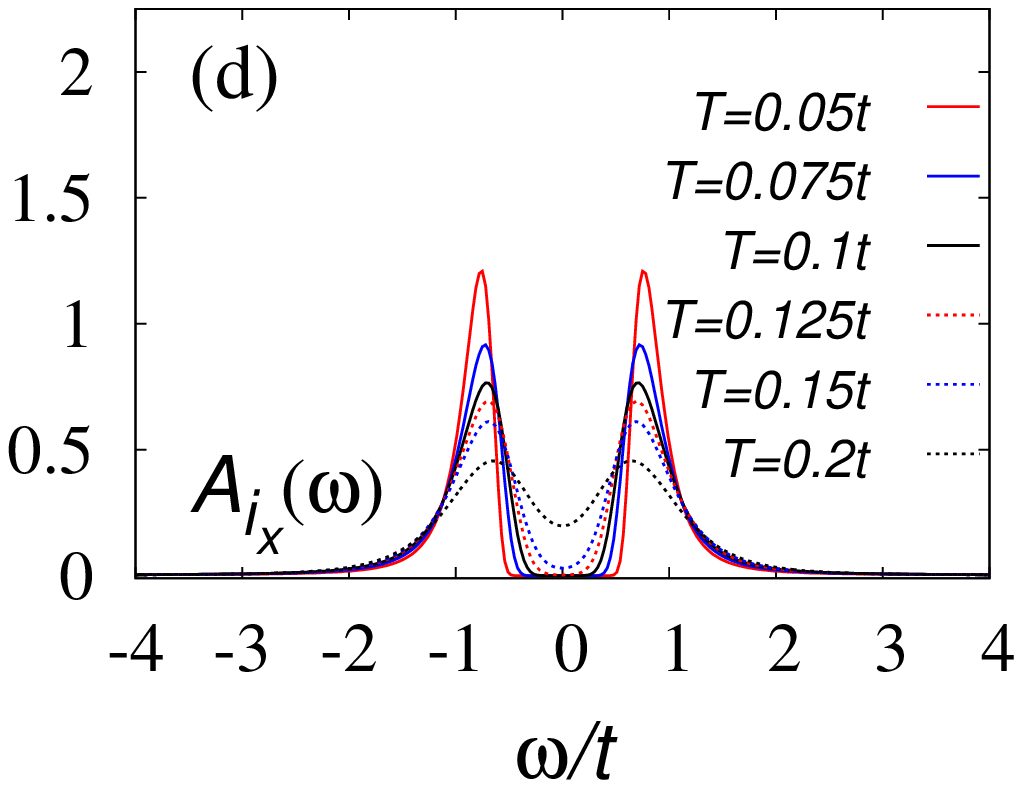}
\end{center}
\end{minipage}
\begin{minipage}{0.45\hsize}
\begin{center}
\includegraphics[width=50mm,clip]{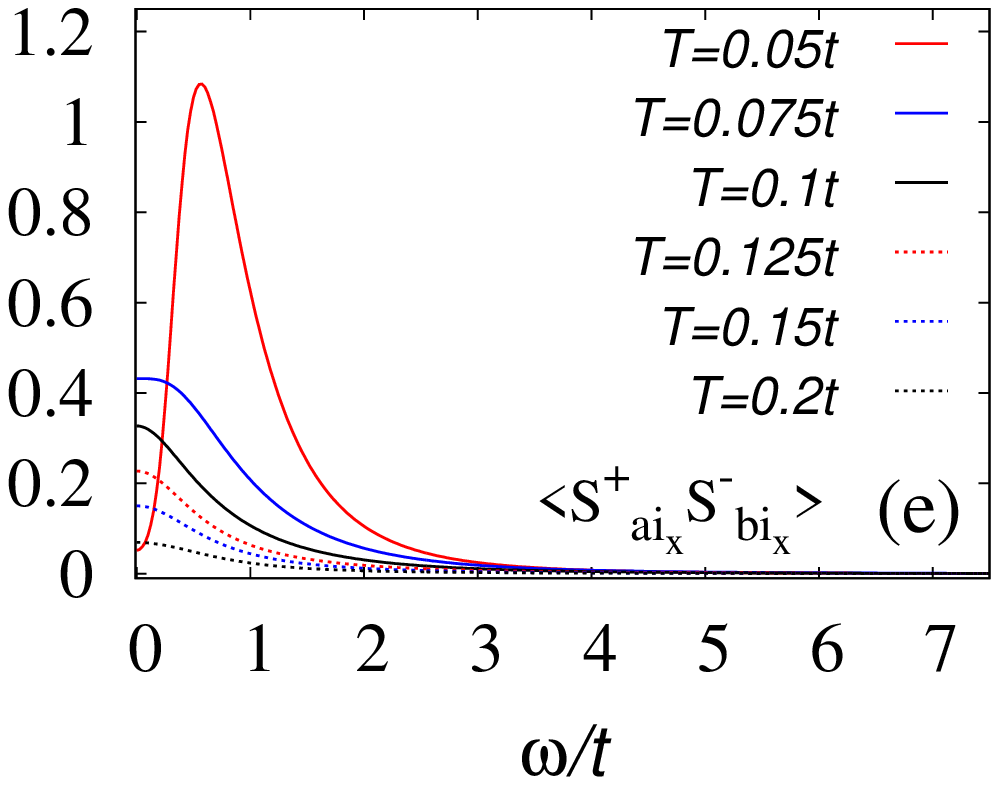}
\end{center}
\end{minipage}
\begin{minipage}{0.45\hsize}
\begin{center}
\includegraphics[width=50mm,clip]{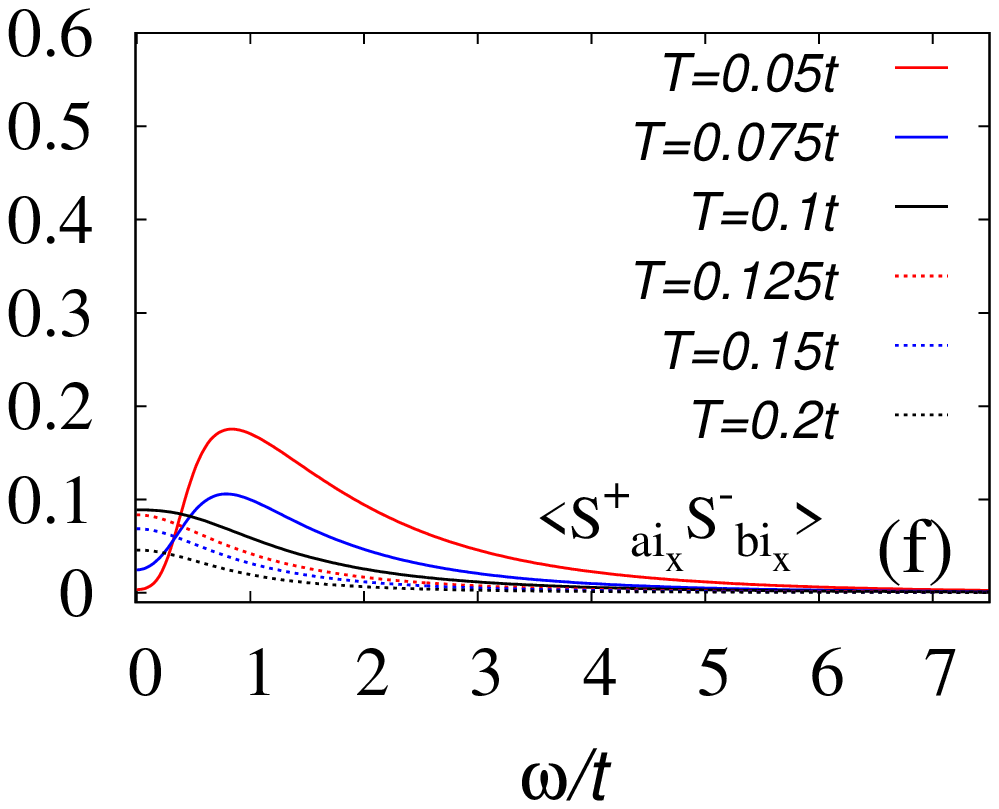}
\end{center}
\end{minipage}
\caption{(Color Online). 
 (a) [(b)]: Color plot of spectral weight of LDOS [$\langle S^+_{ai_x}S^-_{bi_x}\rangle$] at $\omega =0$ and $i_x=0$.
The solid lines denote crossover temperature where spin excitations become gapless. The dashed lines denote crossover temperature where single-particle excitations become gapless.
In the region sandwiched by green lines, the topological edge Mott behaviors are observed. 
(c) and (d): LDOS at $i_x=0$ for several values of temperature. 
(e) and (f): spectral weight of spin excitations ($\langle S^+_{ai_x=0}S^-_{bi_x=0}\rangle$) at $i_x=0$ for several values of temperature.
In panels (c) and (e) [(d) and (f)], data for $(U,J)=(t,0.05t)$ [$(t,0.5t)$] are plotted.
}
\label{fig:spectrum Ueq1}
\end{figure}

To show this, we compute LDOSs and local spin excitations at the edge for $U=t$. 
In Figs.~\ref{fig:spectrum Ueq1} (a) and (b), the spectral weight of the LDOS $A_{i_x}$, and spin excitations $\langle S^+_{i_xa}S^+_{i_xb}\rangle$ for $\omega=0$ at the edge ($i_x=0$) are plotted. 
With increasing temperature, the spectral weights at $\omega=0$ increases, which indicates the restoration of gapless edge modes. Here, the dashed (solid) lines denote crossover temperature where the single-particle excitations (the spin excitations) become gapless, respectively. 
Namely, in the region sandwiched by dashed and solid green lines, the system shows gapless modes only in spin (i.e., bosonic) excitations. Further increasing interactions we can observe the gapless fermionic excitations.
Here, we show the details of the data. The LDOS (the local spin excitation) is plotted in Figs.~\ref{fig:spectrum Ueq1} (c) and (d) [(e) and (f)].
The data for $U=t$ and $J=0.05t$, [see Figs.~\ref{fig:spectrum Ueq1} (c) and (e)] indicate that the spin excitations show a coherence peak for $T=0.075t$ while electron excitations are still gapped at this temperature.
Absence of gapless modes in the low temperature region is more clearly observed for $U=t$ and $J=0.5t$, where the restoration of gapless modes is also observed [see Figs.~\ref{fig:spectrum Ueq1} (d) and (f)].
Performing similar calculations we end up with Figs.~\ref{fig:spectrum Ueq1} (a) and (b) where we can find the emergence of the topological edge Mott state only at finite temperatures.

\section{summary}\label{sec. summary}
In this paper, we have analyzed the reduction of topological classification in the two-dimensional correlated systems by using the R-DMFT+CTQMC.
In particular, we have elucidated the following behaviors.
(i) The topological invariant~(\ref{eq: nu}) takes a nontrivial value even when the reduction occurs. 
(ii) The absence of zero modes becomes consistent with the finite winding number due to the emergence of the zeros of the Green's function only at the edge. The system shows the Mott behaviors only around the edge, resulting in the singularity of the self-energy only at the edge. 
(iii) Increasing temperature removes the zeros of the Green's function emerging only around the edges and restores gapless electron excitations around the edge. Besides, we have found that the temperature effects can change statistical properties of the restored gapless edge modes.

The result (i) indicates that calculating the winding number in Eq.~(\ref{eq: nu}) is insufficient to describe the reduction in the bulk, although it is well-defined even in the presence of electron correlation.
For the direct observation of the reduction in the bulk, it is recently proposed to calculate the angle of partition functions on unoriented space-time.\cite{Shapourian2016,Shiozaki2016} 
The numerical simulation of it is left for our future work.

\section{acknowledgements}
This work is partly supported by a Grand-in-Aid for Scientific Research on Innovative Areas (JSPS KAKENHI Grant No. JP15H05855) and also JSPS KAKENHI (No. 16K05501). 
The numerical calculations were per- formed on supercomputer at the ISSP in the University of Tokyo, and the SR16000 at YITP in Kyoto University.


\begin{thebibliography}{45}
\bibitem [{foo()}]{footnote_first}%
  \BibitemOpen
  \href@noop {} {}\bibinfo {note} {{ In two-dimensional systems, spatial
  fluctuations is important. In our approach, the spatial fluctuation is
  neglected. Hence, to obtain accurate results concerning the order of the
  transition, one should perform simulations by taking into account spatial
  fluctuation }}\BibitemShut {NoStop}%
\end{thebibliography}

%

\end{document}